\documentclass[a4paper,11pt]{article}
\pdfoutput=1 

\usepackage{jcappub} 

\usepackage[T1]{fontenc} 
\usepackage{verbatim}
\usepackage{graphicx}
\usepackage{gensymb}
\usepackage{color}
\usepackage{amssymb}
\usepackage{wasysym}
\usepackage{amsmath}
\usepackage{units}
\usepackage[utf8]{inputenc}
\usepackage{bm}

\usepackage{float}
\usepackage[caption = false]{subfig}

\graphicspath{ {Figures/} }

\title{\boldmath Constraining the Local Burst Rate Density of Primordial Black Holes with HAWC}

\author[a]{A.~Albert}
\author[b]{R.~Alfaro}
\author[c]{C.~Alvarez}
\author[d]{J.C.~Arteaga-Velázquez}
\author[e]{K.P.~Arunbabu}
\author[b]{D.~Avila Rojas}
\author[f]{H.A.~Ayala Solares}
\author[g]{V.~Baghmanyan}
\author[b]{E.~Belmont-Moreno}
\author[h]{S.Y.~BenZvi}
\author[i]{C.~Brisbois}
\author[c]{K.S.~Caballero-Mora}
\author[j]{T.~Capistrán}
\author[j]{A.~Carramiñana}
\author[g]{S.~Casanova}
\author[d]{U.~Cotti}
\author[k]{J.~Cotzomi}
\author[l,m]{E.~De la Fuente}
\author[d]{C.~de León}
\author[a]{B.L.~Dingus}
\author[n]{M.A.~DuVernois}
\author[l]{J.C.~Díaz-Vélez}
\author[i]{R.W.~Ellsworth}
\author[i,1]{K.L.~Engel,\note{Corresponding author.}}
\author[b]{C.~Espinoza}
\author[o]{H.~Fleischhack}
\author[p]{N.~Fraija}
\author[p]{A.~Galván-Gámez}
\author[b]{J.A.~García-González}
\author[p]{F.~Garfias}
\author[p]{M.M.~González}
\author[i]{J.A.~Goodman}
\author[a]{J.P.~Harding}
\author[b]{S.~Hernandez}
\author[o]{B.~Hona}
\author[o]{D.~Huang}
\author[c]{F.~Hueyotl-Zahuantitla}
\author[o]{P.~Hüntemeyer}
\author[p]{A.~Iriarte}
\author[q]{V.~Joshi}
\author[e]{A.~Lara}
\author[p]{W.H.~Lee}
\author[b]{H.~León Vargas}
\author[r]{J.T.~Linnemann}
\author[j]{A.L.~Longinotti}
\author[s]{G.~Luis-Raya}
\author[r]{J.~Lundeen}
\author[t]{R.~López-Coto}
\author[a]{K.~Malone}
\author[r]{S.S.~Marinelli}
\author[k]{O.~Martinez}
\author[i]{I.~Martinez-Castellanos}
\author[u]{J.~Martínez-Castro}
\author[v]{H.~Mart\'inez-Huerta}
\author[w]{J.A.~Matthews}
\author[x]{P.~Miranda-Romagnoli}
\author[d]{J.A.~Morales-Soto}
\author[k]{E.~Moreno}
\author[f]{M.~Mostafá}
\author[g]{A.~Nayerhoda}
\author[y]{L.~Nellen}
\author[z]{M.~Newbold}
\author[r]{M.U.~Nisa}
\author[x]{R.~Noriega-Papaqui}
\author[r,1]{A.~Peisker}
\author[s]{E.G.~Pérez-Pérez}
\author[h]{C.D.~Rho}
\author[i]{C.~Rivière}
\author[j]{D.~Rosa-González}
\author[f]{M.~Rosenberg}
\author[k]{H.~Salazar}
\author[g]{F.~Salesa Greus}
\author[b]{A.~Sandoval}
\author[i]{M.~Schneider}
\author[a]{G.~Sinnis}
\author[i]{A.J.~Smith}
\author[z]{R.W.~Springer}
\author[i]{E.~Tabachnick}
\author[f]{M.~Tanner}
\author[s]{O.~Tibolla}
\author[r]{K.~Tollefson}
\author[j]{I.~Torres}
\author[l,m]{R.~Torres-Escobedo}
\author[n]{T.~Weisgarber}
\author[n]{J.~Wood}
\author[aa]{A.~Zepeda}
\author[a]{H.~Zhou}

\affiliation[a]{Physics Division, Los Alamos National Laboratory, Los Alamos, NM, USA }
\affiliation[b]{Instituto de F\'{i}sica, Universidad Nacional Autónoma de México, Ciudad de Mexico, Mexico }
\affiliation[c]{Universidad Autónoma de Chiapas, Tuxtla Gutiérrez, Chiapas, México}
\affiliation[d]{Universidad Michoacana de San Nicolás de Hidalgo, Morelia, Mexico }
\affiliation[e]{Instituto de Geof\'{i}sica, Universidad Nacional Autónoma de México, Ciudad de Mexico, Mexico }
\affiliation[f]{Department of Physics, Pennsylvania State University, University Park, PA, USA }
\affiliation[g]{Institute of Nuclear Physics Polish Academy of Sciences, PL-31342 IFJ-PAN, Krakow, Poland }
\affiliation[h]{Department of Physics \& Astronomy, University of Rochester, Rochester, NY , USA }
\affiliation[i]{Department of Physics, University of Maryland, College Park, MD, USA }
\affiliation[j]{Instituto Nacional de Astrof\'{i}sica, Óptica y Electrónica, Puebla, Mexico }
\affiliation[k]{Facultad de Ciencias F\'{i}sico Matemáticas, Benemérita Universidad Autónoma de Puebla, Puebla, Mexico }
\affiliation[l]{Departamento de F\'{i}sica, Centro Universitario de Ciencias Exactase Ingenierias, Universidad de Guadalajara, Guadalajara, Mexico }
\affiliation[m]{Department of Physics \& Astronomy, Texas Tech University, USA}
\affiliation[n]{Department of Physics, University of Wisconsin-Madison, Madison, WI, USA }
\affiliation[o]{Department of Physics, Michigan Technological University, Houghton, MI, USA }
\affiliation[p]{Instituto de Astronom\'{i}a, Universidad Nacional Autónoma de México, Ciudad de Mexico, Mexico }
\affiliation[q]{Erlangen Centre for Astroparticle Physics, Friedrich-Alexander-Universit\"{a}t Erlangen-N\"{u}rnberg, Erlangen, Germany}
\affiliation[r]{Department of Physics and Astronomy, Michigan State University, East Lansing, MI, USA }
\affiliation[s]{Universidad Politecnica de Pachuca, Pachuca, Hgo, Mexico }
\affiliation[t]{INFN and Universita di Padova, via Marzolo 8, I-35131,Padova,Italy}
\affiliation[u]{Centro de Investigaci\'on en Computaci\'on, Instituto Polit\'ecnico Nacional, M\'exico City, M\'exico.}
\affiliation[v]{Instituto de F\'isica de S\~ao Carlos, Universidade de S\~ao Paulo, S\~ao Carlos, SP, Brasil}
\affiliation[w]{Dept of Physics and Astronomy, University of New Mexico, Albuquerque, NM, USA }
\affiliation[x]{Universidad Autónoma del Estado de Hidalgo, Pachuca, Mexico }
\affiliation[y]{Instituto de Ciencias Nucleares, Universidad Nacional Autónoma de Mexico, Ciudad de Mexico, Mexico }
\affiliation[z]{Department of Physics and Astronomy, University of Utah, Salt Lake City, UT, USA }
\affiliation[aa]{Physics Department, Centro de Investigacion y de Estudios Avanzados del IPN, Mexico City, DF, Mexico }

\emailAdd{klengel@umd.edu}
\emailAdd{peiskera@msu.edu}

\abstract{Primordial Black Holes (PBHs) may have been created by density fluctuations in the early Universe and could be as massive as $> 10^9$ solar masses or as small as the Planck mass. It has been postulated that a black hole has a temperature inversely-proportional to its mass and will thermally emit all species of fundamental particles via Hawking Radiation. PBHs with initial masses of $\sim 5 \times 10^{14}$ g (approximately one gigaton) should be expiring today with bursts of high-energy gamma radiation in the GeV--TeV energy range. The High Altitude Water Cherenkov (HAWC) Observatory is sensitive to gamma rays with energies of $\sim$300 GeV to past 100 TeV, which corresponds to the high end of the PBH burst spectrum. With its large instantaneous field-of-view of $\sim 2$ sr and a duty cycle over 95\%, the HAWC Observatory is well suited to perform an all-sky search for PBH bursts. We conducted a search using 959 days of HAWC data and exclude the local PBH burst rate density above $3400~\mathrm{pc^{-3}~yr^{-1}}$ at 99\% confidence, the strongest limit on the local PBH burst rate density from any existing electromagnetic measurement.}

\begin{document}
\maketitle
\flushbottom

\section{Introduction}
While there are no known processes in the current Universe that can create black holes with masses less than $\sim 1~ M_{\astrosun}$, conditions in the early Universe were conducive to the formation of black holes with a wide range of masses \cite{Carr:2009jm}. These black holes, with masses ranging from the Planck mass to supermassive black holes ($>10^{9}~M_{\astrosun}$), are called Primordial Black Holes (PBHs). PBH production in the early Universe would have broad observable consequences spanning the largest distance scales (including influencing the development of large-scale structure in the Universe and the primordial power spectrum \cite{Silk2000, Carr:2019yxo, Emami:2017fiy}), to the smallest scales (including enhancing local dark matter clustering \cite{Clesse:2016vqa, Belotsky:2018wph}). In the present Universe, PBHs in certain mass ranges may constitute a non-negligible fraction of dark matter \cite{Carr:2009jm, Belotsky:2014kca, Carr:2016drx}. Since the existence of stellar-mass black holes was recently confirmed during the first observational run of Advanced LIGO \cite{Abbott2017}, there has been a resurgence in support for a PBH component of the total dark matter energy density (e.g., Refs.~\cite{Garc_a_Bellido_2017, Tada:2019amh, Carr:2019yxo}). Limits placed thus far indicate that $f(m)$, the fraction of dark matter that is made up of PBHs, is $\lesssim 10\%$ over a range of masses \cite{Carr:2016drx}.

The prediction that a black hole will thermally radiate, or ‘evaporate,’ with a blackbody temperature inversely proportional to its mass was first developed by Hawking in a calculation that convolved quantum field theory, General Relativity, and thermodynamics \cite{Hawking1974}. The emitted radiation consists of all fundamental particles with masses less than approximately the black hole temperature \cite{MacGibbon1990}. For black holes in the stellar mass range and above, Hawking radiation is nearly negligible. However, for lower-mass PBHs, this process dominates their evolution over time \cite{Glicenstein:2013vha}. PBHs with initial masses of $\sim 5 \times 10^{14}~\mathrm{g}$ should be expiring today producing short bursts lasting a few seconds of high-energy gamma radiation in the GeV--TeV energy range \cite{MacGibbon2008, Ukwatta:2015tza}, making their final moments an ideal phenomenon to observe with HAWC. 

Confirmed detection of a PBH burst---beyond proving their existence and allowing the determination of their relic density and rate-density of evaporation---would provide valuable insights into many areas of physics, including fundamental processes in the very early Universe and particle physics at energies higher than currently achievable by terrestrial accelerators \cite{Khlopov:2008qy}. Even the non-detection of PBH burst events in dedicated searches would yield important constraints about the early Universe \cite{Carr:2009jm}. One of the most important reasons to search for PBHs is to constrain the cosmological density fluctuation spectrum in the early Universe on scales smaller than those constrained by the cosmic microwave background \cite{Ukwatta2016}. A particularly interesting question is whether or not PBHs were formed from the quantum fluctuations associated with many different types of inflationary scenarios \cite{Carr:2009jm}. Detection or upper limits on the number density of PBHs can thus also inform inflationary models.

Numerous detectors have searched for PBH burst events using both direct (e.g., \cite{Archambault:2017asc, Glicenstein:2013vha, Fermi-LAT:2018pfs}) and indirect (e.g., \cite{Carr:2009jm, Wright:1995bi, Abe2011}) methods. These methods explore the PBH distribution at various distance scales, from cosmological scales down to within a parsec. At cosmological distances, the PBH density can be probed using the $100$ MeV extragalactic gamma-ray background \cite{Carr:2009jm, Hawking1974, MacGibbon2008, Carr:2005zd, Gehrels2012, Halzen:1991uw, Page1976}. On the Galactic scale, the local PBH density and anisotropy can be studied using the $100$ MeV gamma-ray measurements \cite{Wright:1995bi}; on kiloparsec scales, a PBH burst limit can be placed using the Galactic antiproton background \cite{Abe2011}; and on the parsec scale, the PBH burst limits can be set directly by searching for the detection of individual evaporating PBHs \cite{Abdo:2014apa, Archambault:2017asc, Glicenstein:2013vha, Fermi-LAT:2018pfs}. In this work (an extension of the analysis presented at the 36th International Cosmic Ray Conference in Madison, WI, USA \cite{Engel2019}), the High Altitude Water Cherenkov (HAWC) Observatory searches for PBH bursts at the parsec scale.

This paper is structured as follows. The HAWC Observatory is described in Section \ref{sec:hawc}. Section \ref{sec:theory} provides a description of the model of PBH bursts. The data collection and analysis procedure is presented in Section \ref{sec:analysis}. Section \ref{sec:disc} provides a discussion of the statistical and systematic uncertainties and concludes the paper.
\section{HAWC Observatory}\label{sec:hawc}

The HAWC Observatory, a successor to the Milagro Observatory \cite{Atkins2000}, is a very-high-energy (VHE) ground-based air shower array located on the side of the Sierra Negra volcano in Mexico at an altitude of 4,100 m above sea level. It has a wide field-of-view of $\sim2$ sr and an operational energy range of $\sim$ 300 GeV--100 TeV. HAWC consists of 300 cylindrical water tanks in the main array covering a total area of 22,000 m$^{2}$. Each tank in the main array is 7.3 m in diameter and 4.5 m deep, and is equipped with four (three $8''$ and one $10''$ in diameter) upward-facing photomultiplier tubes (PMTs) anchored to the bottom of the tank. HAWC has completed installing and integrating an additional ``outrigger'' array \cite{Joshi:2019ycc} composed of 345 cylindrical tanks 1.55~m in diameter and 1.65~m deep, each containing a single $8''$ PMT. The outriggers are arranged in a concentric, circular, symmetric pattern around the main array, covering an additional instrumented area of $\sim4.5$ times that of the main array. However, the analysis presented in this work includes only data from the main HAWC array.  

In both the main array and the outriggers, the PMTs detect Cherenkov light from secondary particles created in extensive air showers induced by VHE gamma rays incident on Earth's atmosphere. The main data acquisition system measures the arrival time of secondary particles on the ground and the amplitude of the PMT signals. This information is used to reconstruct the arrival direction and energy of the primary particle. The angular resolution of HAWC ranges from $\sim0.2\degree$ (68\% containment) to $~1.0\degree$, depending on the fraction of PMTs hit by the resulting shower \cite{Abeysekara:2017mjj}. With these features and a high duty cycle of greater than 95\%, HAWC is ideally suited to continuously monitor the Northern Hemisphere sky for high-energy emission from gamma-ray transients such as PBHs.


HAWC's wide field-of-view and continuous operations are advantageous for detecting burst transients with emission durations shorter than the slewing times of Imaging Atmospheric Cherenkov telescopes (IACTs). These features are also key for this analysis as the sensitivity to a PBH burst rate density is determined by the total observable volume and the exposure time.
\section{Theory}\label{sec:theory}
\subsection{Primordial Black Hole Burst Spectrum}\label{sec:spectrum}
As a PBH radiates, it continually loses mass and its temperature increases to very high energies \cite{Hawking1974}. The manner in which the PBH expires depends on the physics at this energy scale. The chosen high-energy particle physics model determines the energy spectrum. In this work, as in Ref. \cite{Abdo:2014apa}, we assume the Standard Evaporation Model (SEM) \cite{MacGibbon1990, MacGibbon1991} as our emission and particle physics model. In the SEM---based on the Standard Model of particle physics---a PBH should directly radiate those fundamental particles whose wavelengths (Compton wavelength $\lambda_{c} = h/mc$ for massive particles) are comparable to the size of the black hole. When the black hole temperature exceeds the Quantum Chromodynamics confinement scale of $\sim$250--300 MeV, quarks and gluons should be radiated \cite{MacGibbon1990, Page1976, MacGibbon:2015mya}. On astrophysical timescales, the final products will decay to the lightest Standard Model particles: photons, neutrinos, electrons, positrons, protons, and anti-protons \cite{MacGibbon1990}.

\begin{figure}[t] 
 \centering
 \includegraphics[width=0.7\linewidth]{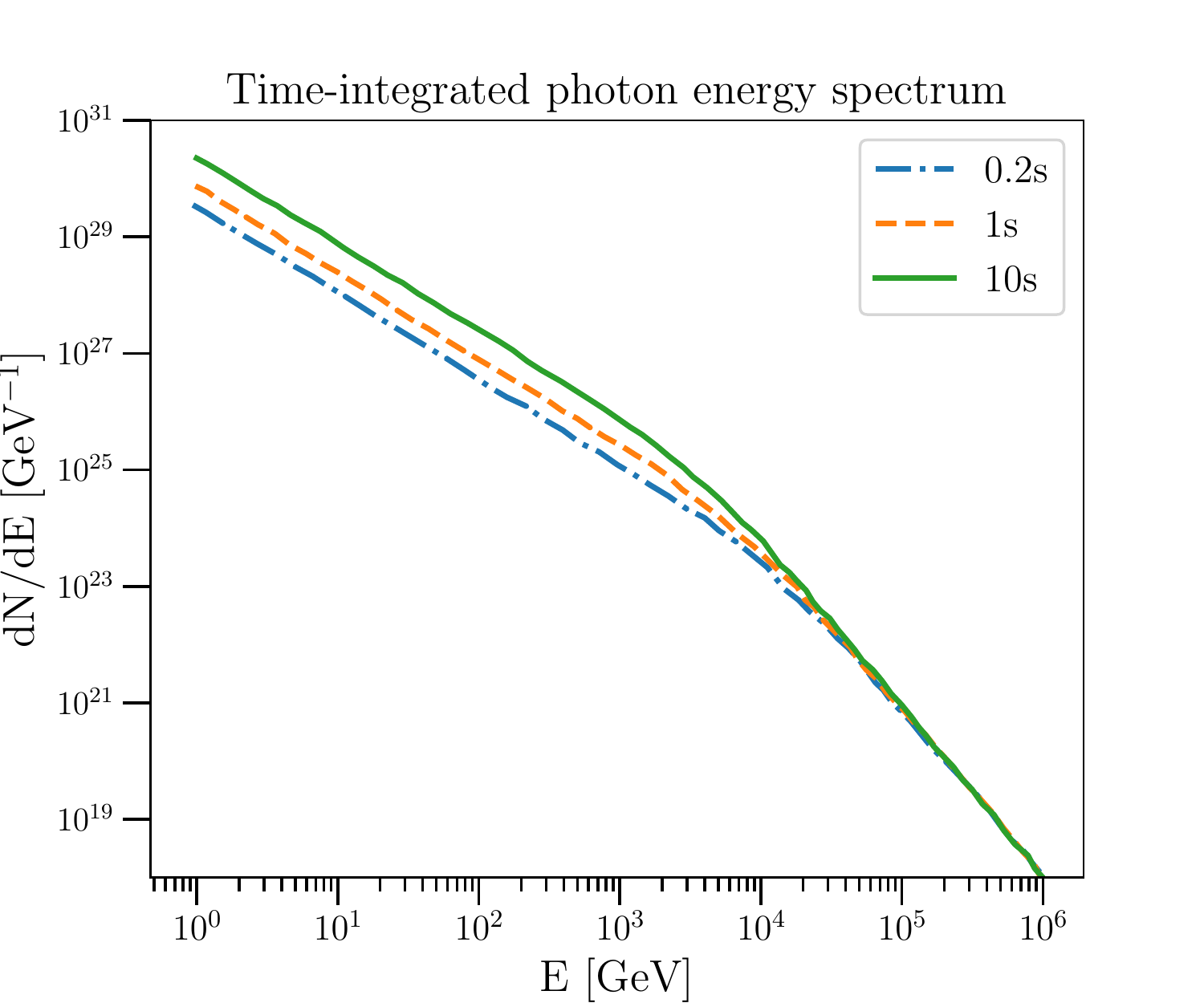}
 \caption{\label{fig:spectrum} The photon spectrum, integrated over the final black hole evaporation lifetime intervals $\tau$ = 0.2, 1, and 10 seconds. Adapted from Ref. \cite{Ukwatta2016}.}
 \end{figure}

In the SEM, the black hole temperature ($T$) can be expressed in terms of the remaining lifetime ($\tau$) of the black hole (that is, the time left until the black hole finishes evaporating) as \cite{MacGibbon1991, Petkov2008},
\begin{equation}\label{BHTemp}
    T \simeq 7.8\times 10^3 \left(\frac{1~ \mathrm{s}}{\tau}\right)^{\nicefrac{1}{3}} \enspace \mathrm{GeV}\enspace,
\end{equation}
for $T \ll T_p$, where $T_p$ is the Planck temperature ($1.22 \times 10^{19}$ GeV). Note that here we are using units such that the Boltzmann constant $k=1$. The emission rate increases as the black hole shrinks \cite{Page1976}. For black holes with temperatures greater than several GeV at the start of the observation, the time-integrated photon flux can be parameterized as \cite{Bugaev:2007py, Petkov2008, Ukwatta2016},
\newcommand\mycom[2]{\genfrac{}{}{0pt}{}{#1}{#2}}
\begin{equation}\label{photon-param}
    \frac{dN_{\gamma}}{dE_{\gamma}} \approx 9 \times 10^{35} \begin{cases} \Big(\frac{1~\mathrm{GeV}}{T}\Big)^{\nicefrac{3}{2}} \Big( \frac{1~\mathrm{GeV}}{E_{\gamma}}\Big)^{\nicefrac{3}{2}} \makebox[0.4cm]{} \mathrm{GeV^{-1}}\enspace & \mathrm{for} ~E_{\gamma} < T \\
    \enspace\enspace\enspace\enspace\enspace\enspace \Big(\frac{1~\mathrm{GeV}}{E_{\gamma}}\Big)^{3} \makebox[0.4cm]{} \mathrm{GeV^{-1}}\enspace & \mathrm{for} ~E_{\gamma} \geq T
    \end{cases}\enspace,
\end{equation}
for gamma-ray energies E $\gtrsim 10$ GeV. This parameterization includes both directly radiated photons and those produced by the decay of other directly radiated species. Figure \ref{fig:spectrum} shows the total gamma-ray spectrum for the PBH remaining lifetimes ($\tau$) examined in this work, $\tau=$  0.2~s, 1~s, and 10~s. 

\subsection{Creating the Model}\label{sec:model}
The first step in modeling PBH bursts is to simulate burst source points in HAWC's field-of-view. We used a Monte Carlo simulation to generate a random set of PBH burst events assuming a local burst rate density, $\dot\rho$, of $10^4~\mathrm{pc^{-3}~yr^{-1}}$. We chose this burst rate density merely because it is close to HAWC's predicted sensitivity \cite{Abdo:2014apa}, but it could have been set to be any positive non-zero value. The events were distributed uniformly in a $50^{\degree}$ cone centered in the HAWC field-of-view, out to a distance of 0.5 pc. Beyond 0.5~pc, even PBH bursts simulated for 10~s at HAWC's zenith did not produce an observable signal within HAWC. 

The parameterization of the time-integrated photon flux given in eq.~(\ref{photon-param}) can be used to calculate the expected number of photons detectable by an observatory on the Earth's surface. For a PBH burst of duration $\tau$ at a non-cosmological distance $r$ and zenith angle $\theta$, the number of expected photons may be expressed as,
\begin{equation}\label{exp-photon}
    \mu(r, \theta, \tau) = \frac{(1 - f)}{4 \pi r^{2}} \int_{E_1}^{E_2} \frac{dN(\tau)}{dE} A(E, \theta) dE\enspace,
\end{equation}
where $dN/dE$ is the PBH gamma-ray spectrum integrated from remaining lifetime $\tau$ to 0. The energies $E_1$ and $E_2$ correspond to the energy range of the detector, $A(E, \theta)$ is the effective area of the detector as a function of photon energy and zenith angle, and $f$ is the dead time fraction of the detector. While eq. (\ref{exp-photon}) schematically illustrates how to approach this portion of the analysis, the actual calculations used a forward-folding approach \cite{Younk2016}---described below---to better characterize the detector.

\begin{figure}[thb!] 
 \centering
 \includegraphics[width=0.55\linewidth]{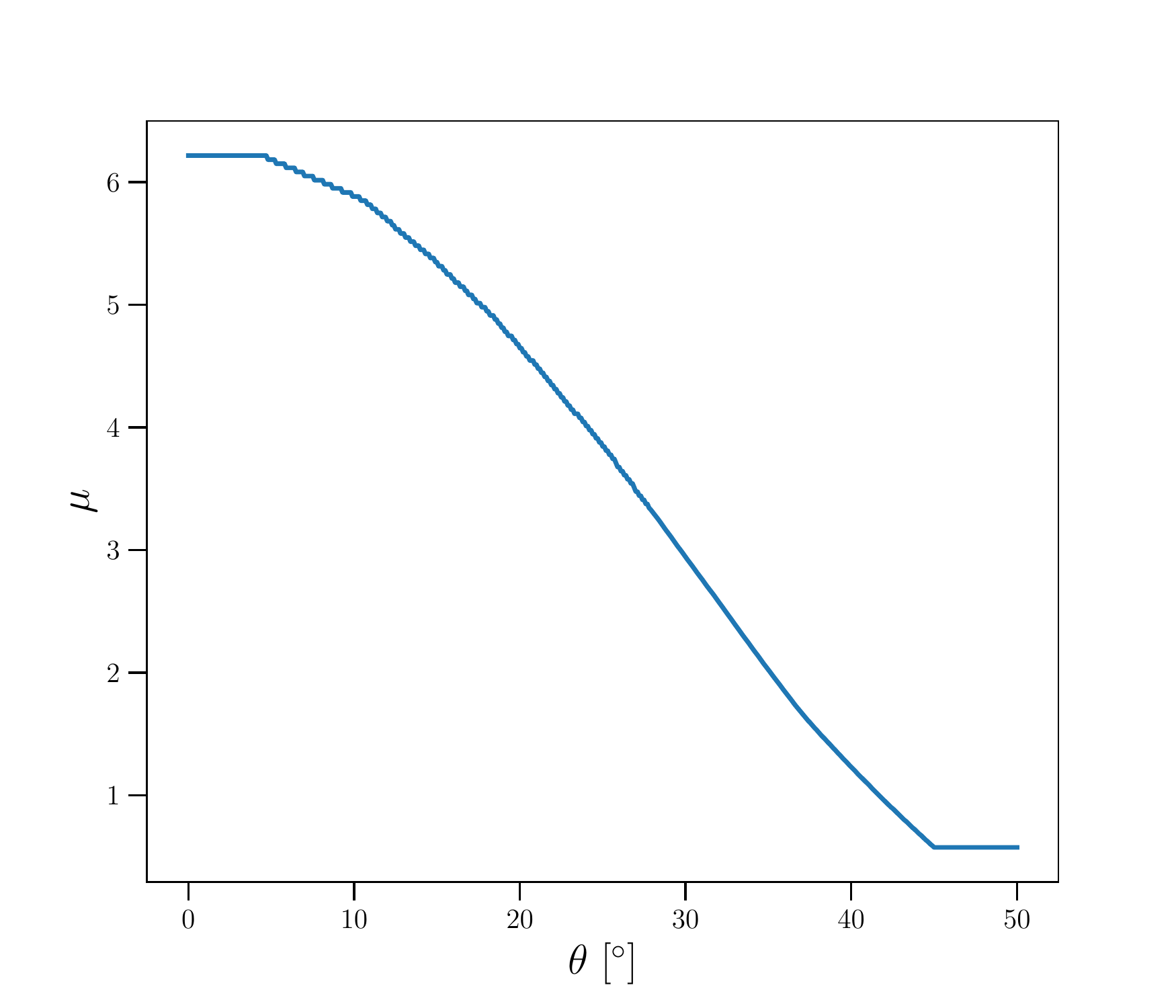}
 \caption{\label{fig:zebra} Sample expected signal at HAWC ($\mu$; see eq. (\ref{exp-photon})) from fixed zenith angles ($\theta$) out of HAWC's instrument response code, ZEBRA \cite{IsraelThesis}, for the PBH spectrum parameterized by eq. (\ref{photon-param}). The sample PBH is assumed to have a remaining lifetime of 0.2~s and be located at a distance of 0.05~pc. The slope plateaus above $45^{\degree}$ due to HAWC's low sensitivity at those zenith angles.}
 \end{figure}

To determine the expected signal at HAWC from each simulated PBH in the field-of-view, we utilized a forward-folding approach based on a functionality within the HAWC software called ZEBRA, which stands for ZEnith Band Response Analysis \cite{IsraelThesis}. ZEBRA uses simulation to characterize the response of the HAWC detector as a function of zenith angle, which is then convolved with the expected spectrum of the source to estimate the counts observed from that source during an arbitrary period of time. Our source model assumed the timescale of the emission was short enough to be considered as all coming from a single zenith angle, with the PBH spectrum following eq.~(\ref{photon-param}) \cite{Ukwatta2016}. This procedure was completed for each of the remaining lifetimes we consider. The number of estimated counts given by ZEBRA for a simulated PBH with remaining lifetime of 0.2~s at a distance of 0.05~pc is shown in Figure \ref{fig:zebra}.

\begin{center}
\begin{figure}
\centering
\subfloat[$\tau=$~0.2~s]{\includegraphics[width= 3.2in]{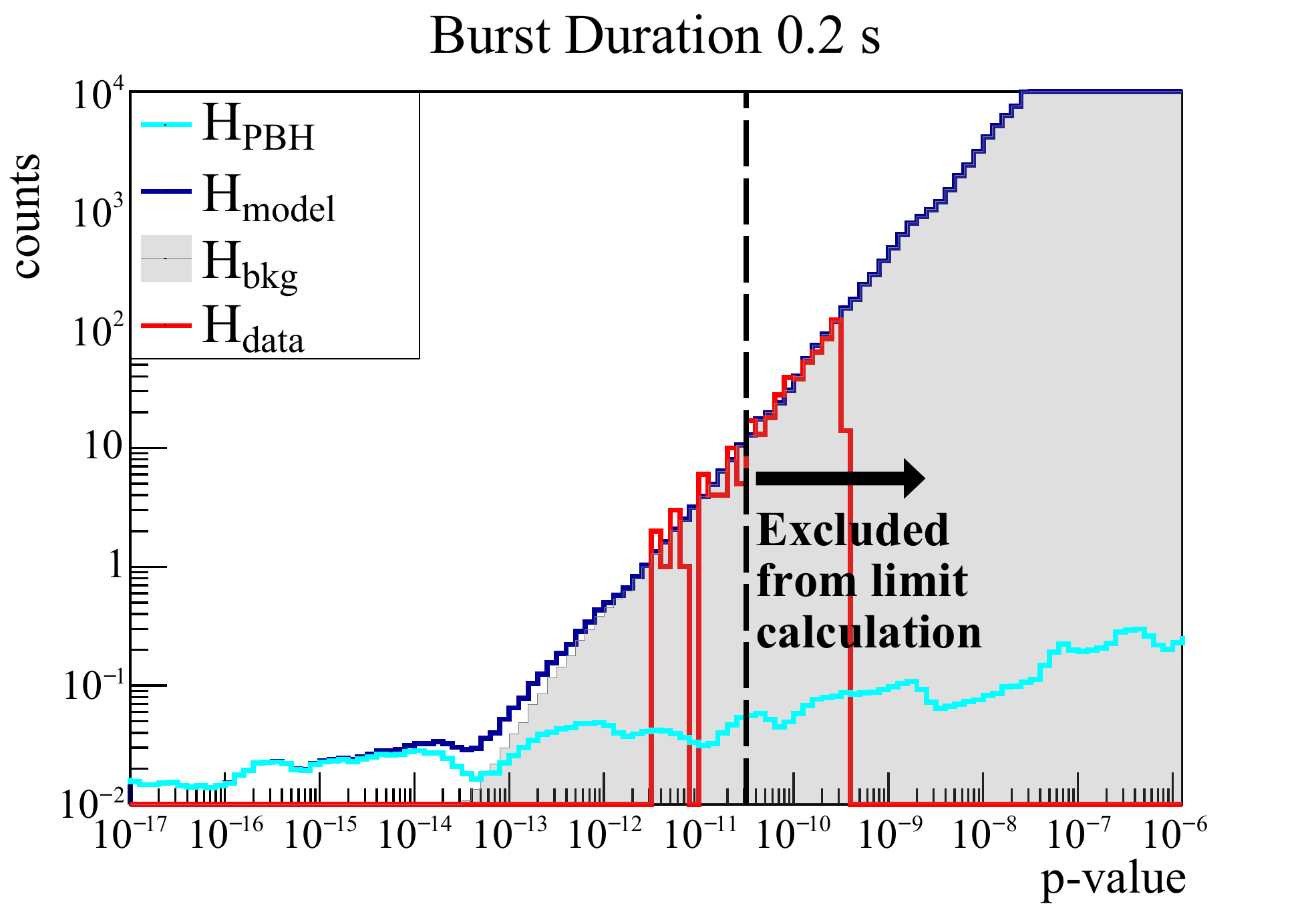}}
\subfloat[$\tau=$~1~s]{\includegraphics[width= 3.2in]{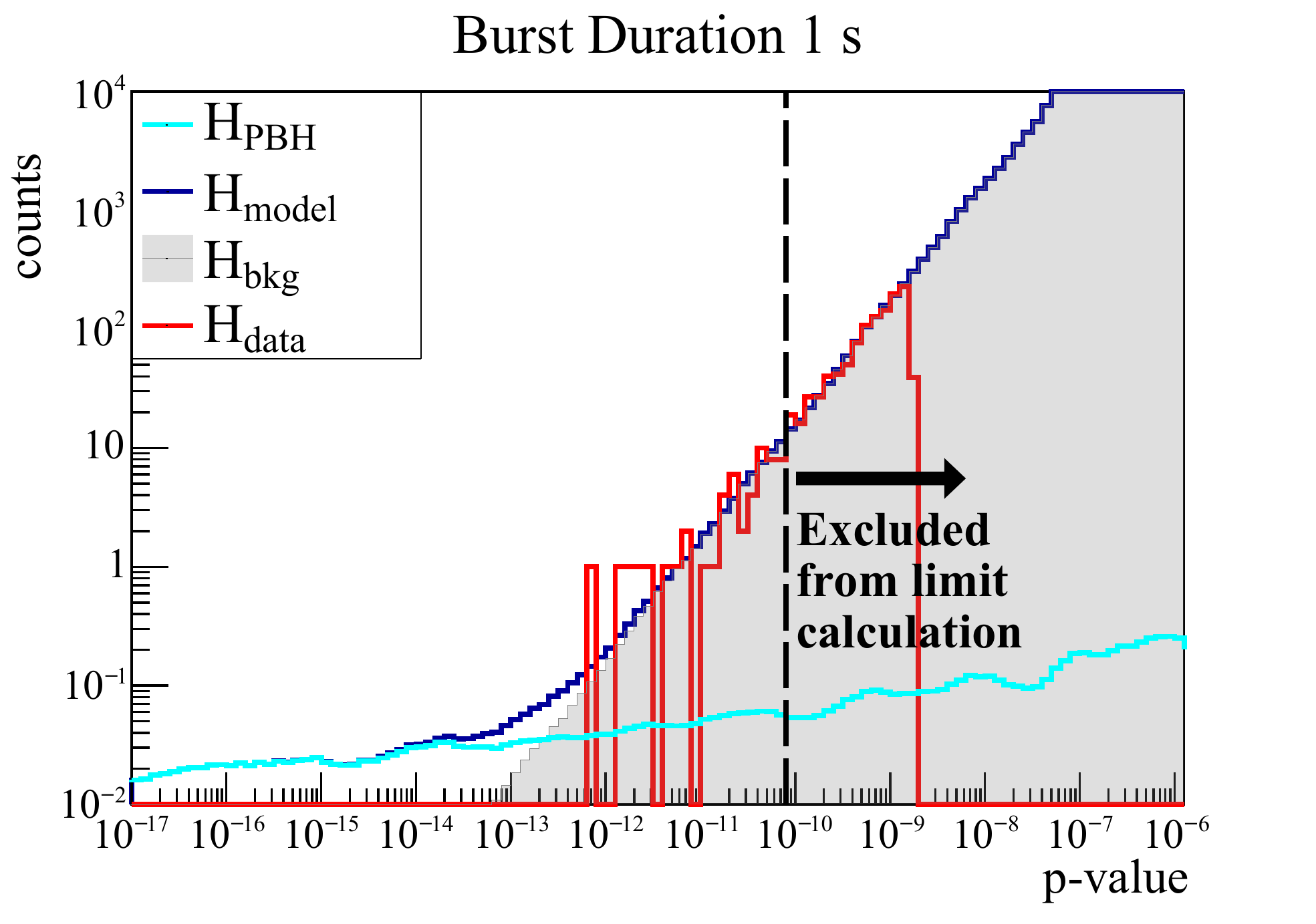}} \\
\subfloat[$\tau=$~10~s]{\includegraphics[width= 3.2in]{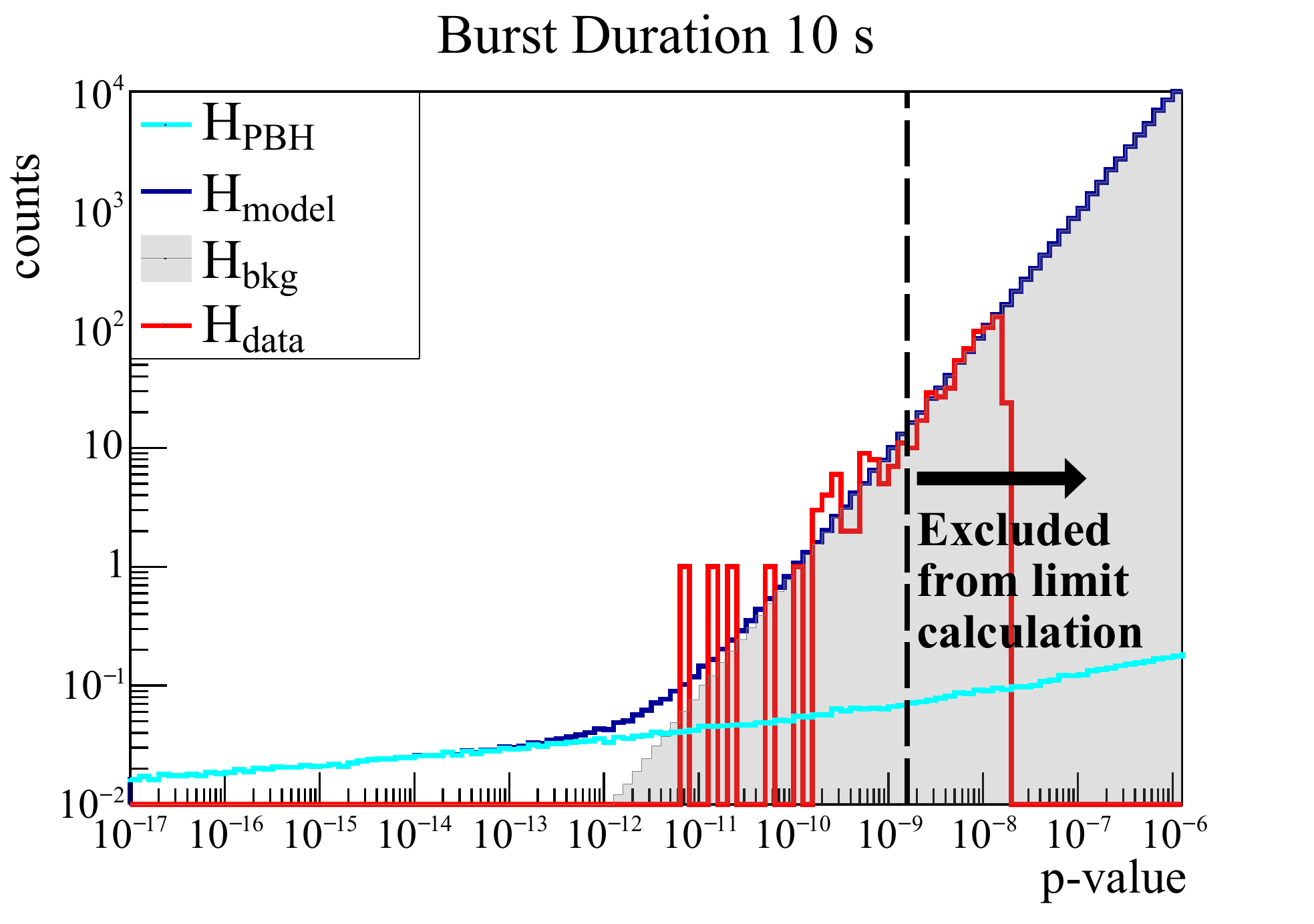}}
\caption{\label{models} Distribution of the p-values of the histograms $H_{\mathrm{PBH}}$ and $H_{\mathrm{model}}$, as well as $H_{\mathrm{bkg}}$ and $H_{\mathrm{data}}$ (described in Sections~\ref{sec:GRBdata} \& \ref{sec:limit}), for all three remaining lifetimes in this search. $H_{\mathrm{PBH}}$ corresponds to a burst rate density of $\dot\rho=10^4~\mathrm{pc^{-3}~yr^{-1}}$. The vertical dashed black line indicates $p_{\mathrm{thr}}$, the passing threshold for data inclusion in this analysis. Note that both the $H_{\mathrm{bkg}}$ \& $H_{\mathrm{PBH}}$ distributions were drawn from a sample much larger than the duration of $H_{\mathrm{data}}$ and were scaled to match this duration, yielding smooth, well-behaved distributions.}
\end{figure}
\end{center}

Using the estimated signal, we calculated the Poisson probability (p-value) of obtaining $N$ or more counts given the background, $B$, for each burst event, 
\begin{equation}\label{p-val}
    p = \mathrm{Pr}(n\geq N) = \sum_{i=N}^{\infty}\frac{B^{i}e^{-B}}{i!} = \frac{\gamma(N,B)}{\Gamma(N)}\enspace,
\end{equation}
where $N = (\mu + B)$ and $\gamma$ is the lower incomplete gamma function. 
These p-values were compiled into a histogram, $H_{\mathrm{PBH}}$ for each lifetime searched: 0.2, 1, and 10~s. Once the signal counts were tabulated in $H_{\mathrm{PBH}}$, we computed a second histogram of counts due to the background of charged cosmic rays. This histogram, denoted $H_{\mathrm{bkg}}$ (described in Sections~\ref{sec:GRBdata} \& \ref{sec:limit}), was added to $H_{\mathrm{PBH}}$ to produce a model of the data: $H_{\mathrm{model}}=H_{\mathrm{bkg}}+H_{\mathrm{PBH}}$, as shown in Figure \ref{models}.
\section{Analysis}\label{sec:analysis}
\subsection{HAWC Transient Search Data}\label{sec:GRBdata}
The data set used for the PBH burst search, $H_{\mathrm{data}}$, consists of data from a self-triggered all-sky transient search conducted from March 2015 through May 2018 \cite{WoodThesis}. This program, originally designed for use in a gamma-ray burst (GRB) analysis, continuously searches for transients at energies above a few hundred GeV with sliding time windows of lengths 0.2, 1, and 10 seconds--- corresponding to typical timescales of peak structures within GRBs. It detects clusters of events above the cosmic-ray rate and stores the probability of observing each cluster of events under a background-only hypothesis. Events were required to have fired at least 70 PMTs and pass a loose hadron-rejection cut optimized for gamma-ray energies $<10$~TeV. The background, used herein to form $H_{\mathrm{bkg}}$, is estimated using a 1.75 hour integration \cite{WoodThesis}. To be computationally efficient, these p-values are saved only if they fall below a certain value. This feature can be seen in the righthand edge of $H_{\mathrm{data}}$ in Figure~\ref{models}. 

The search is performed by shifting each window forward in time and binning air shower events during that window using a grid of $2.1^{\degree} \times 2.1^{\degree}$ square spatial bins covering all points within $50^{\degree}$ of detector zenith. The step size of each shift is less than the bin size to allow for desired overlap. The size of the overlap is optimized to allow for fine tuning on the spatial position of the events while avoiding strong correlations between pixels, and is a different amount for each search duration. Points outside a zenith angle of $50^{\degree}$ are excluded from the spatial search because most photons at the energies expected from a transient burst signal with $\theta>50^{\degree}$ do not have sufficient energy to reach HAWC due to attenuation of air showers at larger atmospheric depth. The fixed-width time windows and square bins were utilized---rather than attempting to fit a light curve profile (which would improve sensitivity)---to make a full search of the HAWC field-of-view computationally tractable \cite{WoodThesis}.

The analysis threshold for each remaining lifetime searched in this PBH analysis was chosen to be the p-value where there are 10 events in the background histogram to ensure the limit placed is conservative and not sensitive to fluctuations in the data. A vertical black line indicates this threshold in Figure~\ref{models}. Above this analysis threshold, where $p$ > $p_\mathrm{thr}$, we have a fiducial region in which we can verify---independently of the analysis---that the background and data are statistically equivalent; that is, to ensure that the overall rate normalization seen in data matches the background estimate from simulation. As we can be sure the data are well-described by a background-only distribution  above $p_{\mathrm{thr}}$, we are confident in our extrapolation of the background-only model to the low-statistics area. 

\subsection{Calculating an Upper Limit}\label{sec:limit}
Finding no statistically significant PBH signal in the data, we computed upper limits on the local burst rate density of PBHs. We began by choosing an approximate value for the local burst rate density of PBHs, $\dot\rho \sim 10^4$ $\mathrm{pc^{-3}~yr^{-1}}$--- the PBH rate thrown in our Monte Carlo simulation as described in Section~\ref{sec:model}. We then scanned over $\dot\rho$ until we found the most probable value of $\dot\rho$ given the data, as well as the 99\% upper limit.

To determine the 99\% limit, we began by calculating the test statistic for $\dot\rho$, 
\begin{equation}\label{TS-basic}
TS = 2[\ln{\mathcal{L}_1} - \ln{\mathcal{L}_0}] = 2[\ln{\mathcal{L}_1}(\dot\rho) - \ln{\mathcal{L}_0}(\dot\rho=0)]\enspace,
\end{equation}
where $\mathcal{L}_0$ is the background Poisson likelihood and $\mathcal{L}_1$ is our model Poisson likelihood. The TS follows a $\chi^2$ distribution with one degree-of-freedom based on Wilks' theorem \cite{Wilks1938}. In general for binned data, the log-likelihood of a Poissonian variable $x$, with mean $\lambda$ for an independently and identically distributed sample of size $n$, is given by,
\begin{equation}\label{LL-basic}
\ln{[\mathcal{L}(\lambda_{tot}|x_1, x_2, ... x_n)]} = \left(\sum_{i=1}^{n} x_i \ln{\lambda_{i}}\right) - n\lambda_{tot} - \ln{\left(\prod_{i=1}^{n} x_i!\right)}\enspace,
\end{equation}
such that we can write, summing over bins, $p$, instead of sample size, 
\begin{equation}\label{LL-bkg}
\ln{(\mathcal{L}_0)} = \sum_{p}\big[H_{\mathrm{\mathrm{data}}}(p)\ln{\big(H_{\mathrm{bkg}}(p)\big)} - H_{\mathrm{bkg}}(p)\big]\enspace,
\end{equation}
where $H_{\mathrm{data}}$ is a histogram of the HAWC data p-values produced by HAWC's transient burst search \cite{Wood2018, WoodThesis} and $H_{\mathrm{bkg}}$ is a histogram of p-values generated by Monte Carlo using HAWC background distributions, also from the transient burst search. Similarly, the model log-likelihood can be written as, 
\begin{equation}\label{LL-model}
\ln{(\mathcal{L}_1)} = \sum_{p}\big[H_{\mathrm{data}}(p)\ln{\big(H_{\mathrm{model}}(p)\big)} - H_{\mathrm{model}}(p)\big]\enspace,
\end{equation}
where $H_{\mathrm{model}}(p) = H_{\mathrm{bkg}}(p) + H_{\mathrm{PBH}}(p)$. Note that both expressions neglect the factorial term in the likelihood eq.~(\ref{LL-basic}) as it will cancel when evaluating eq.~(\ref{TS-basic}).

We then found the value of $\dot\rho$ that yielded the largest $TS$ value from eq.~(\ref{TS-basic}), $TS_{\mathrm{max}}$. Scanning over increasing values of $\dot\rho$, we stopped when the change in $TS$ from the maximum reached $\Delta TS=5.41$, which corresponds to a one-sided 99\% confidence interval. 
Although our strictest limit was placed for a remaining lifetime of 0.2~s, in an effort to place a conservative limit we report the remaining lifetime for which HAWC's sensitivity was predicted to be the strongest--- corresponding to $\tau=10$~s,
\begin{equation}
    \dot\rho < 3400 \substack{+400 \\ -100} ~\mathrm{pc^{-3} yr^{-1}}\enspace,
\end{equation}
the strictest limit yet placed on the local PBH burst rate density. The uncertainties in the limit are systematic only and are described in Section \ref{sec:sys}.
\section{Discussion}\label{sec:disc}

\begin{center}
\begin{figure}[htb!]
\includegraphics[width=1\textwidth]{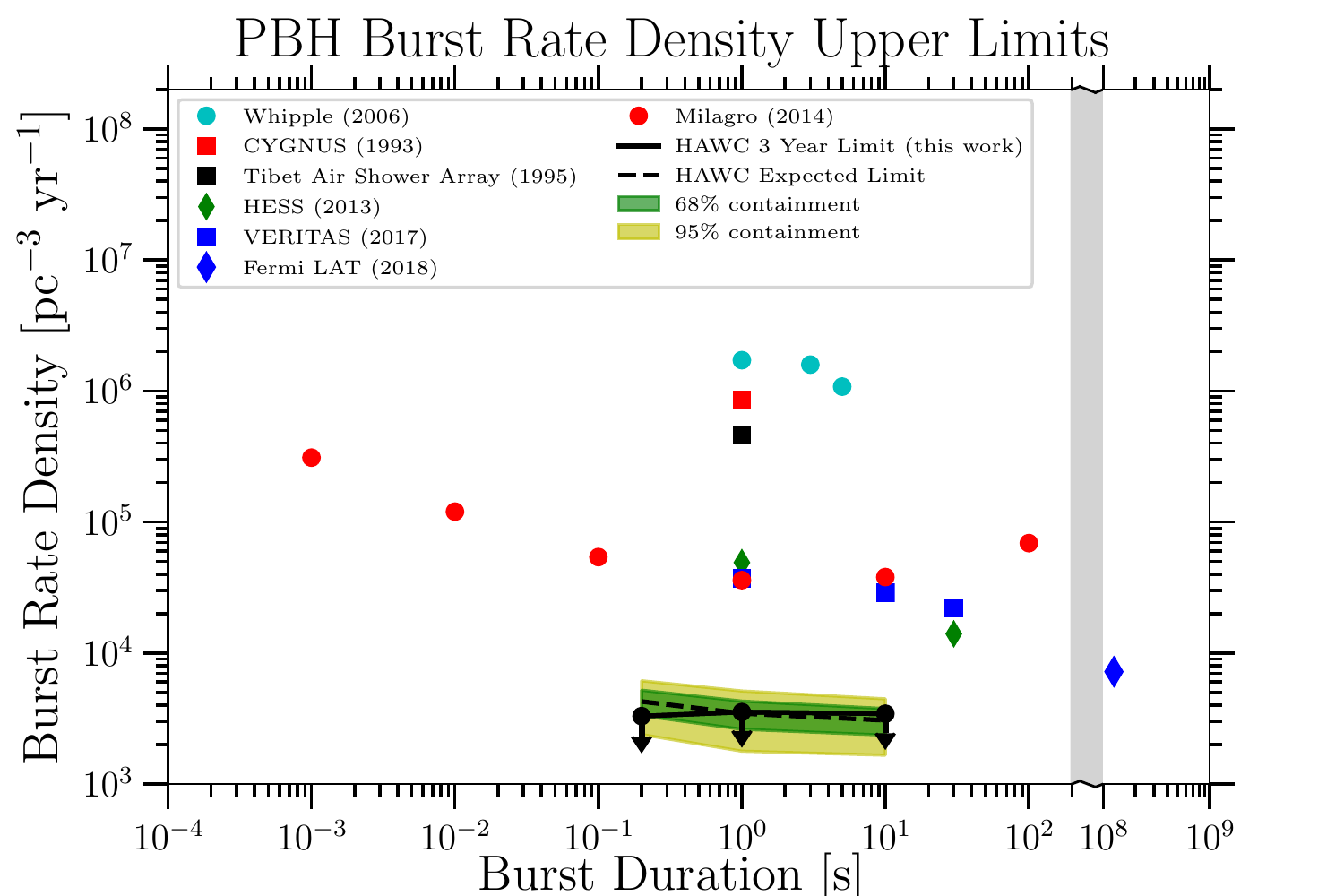}
\caption{Comparison of the HAWC 99\% confidence level upper limits at 0.2, 1, and 10~s with the upper limits from Whipple \cite{Linton2006}, CYGNUS \cite{Alexandreas1993}, the Tibet Air Shower Array \cite{Amenomori1995}, H.E.S.S. \cite{Glicenstein:2013vha}, VERITAS \cite{Archambault:2017asc}, \textit{Fermi}-LAT  \cite{Fermi-LAT:2018pfs}, and Milagro \cite{Abdo:2014apa}. Also shown are the expected limits, 68\% containment, and 95\% containment for the null hypothesis based on HAWC sensitivity. Results displayed between the three explicitly evaluated burst durations are interpolations. While the statistical fluctuations of the limits show possible turn-over in the inferred limit, the sensitivity line indicates a smooth trend toward worse limits for shorter remaining lifetimes.}
\label{fig:result}
\end{figure}
\end{center}

The 99\% upper limits on the PBH burst rate density for each of the three remaining lifetimes searched are listed in Table~\ref{hawc_limits} and shown in Figure~\ref{fig:result}. The expected limits, as well as the 68\% and 95\% containment for the null hypothesis, are also shown in Figure~\ref{fig:result}. The containment bands and expected limits are calculated using 1000 simulations containing no PBHs. Note that the bands are purely statistical and indicate that the upper limits are compatible with the expected sensitivity of HAWC assuming only background events. Our reported limit, corresponding to a remaining lifetime of 10~s, is also presented in Table~\ref{compare} for direct comparison with the results of other direct searches for PBHs. All limits shown in Figure~\ref{fig:result} and Tables~\ref{hawc_limits} \& \ref{compare} were obtained based on the PBH Standard Emission Model \cite{MacGibbon1990,Halzen:1991uw}.
\begin{table}
\begin{center}
\begin{tabular}{ c|c } 
 \textbf{Burst duration} & \textbf{Burst Rate Upper Limit} \\ 
 \hline
 \hline
 $0.2~\mathrm{s}$ & 3300 $\substack{+300 \\ -100}~\mathrm{pc^{-3} yr^{-1}}$ \\
 $1~\mathrm{s}$ & 3500 $\substack{+400 \\ -200}~\mathrm{pc^{-3} yr^{-1}}$ \\
 $10~\mathrm{s}$ & 3400 $\substack{+400 \\ -100}~\mathrm{pc^{-3} yr^{-1}}$ \\
\end{tabular}
\caption{The 99$\%$ upper limits on the PBH burst rate density for the three remaining lifetimes searched. The uncertainties are systematic only, and are described in Section \ref{sec:sys}.}
\label{hawc_limits}
\end{center}
\end{table}
\begin{table}
\begin{center}
\begin{tabular}{ c||c|c|c } 
 \textbf{Experiment} & \textbf{Burst Rate Upper Limit} & \textbf{Search Duration} & \textbf{Reference} \\ 
 \hline
 \hline
  Milagro & 36000 $\mathrm{pc^{-3} yr^{-1}}$ & $1~\mathrm{s}$ & \cite{Abdo:2014apa} \\ 
  VERITAS & 22200 $\mathrm{pc^{-3} yr^{-1}}$ & $30~\mathrm{s}$ & \cite{Archambault:2017asc} \\ 
  H.E.S.S. & 14000 $\mathrm{pc^{-3} yr^{-1}}$ & $30~\mathrm{s}$ & \cite{Glicenstein:2013vha} \\
  Fermi-LAT & 7200 $\mathrm{pc^{-3} yr^{-1}}$ & $1.26 \times 10^8~\mathrm{s}$ & \cite{Fermi-LAT:2018pfs} \\
  \textbf{HAWC 3 yr.} & \textbf{3400} \bm{$\mathrm{pc^{-3} yr^{-1}}$} & \bm{$10~\mathrm{s}$} & \textbf{This Work} \\
\end{tabular}
\caption{The strongest limit on the burst rate density of PBHs for each of the five detectors most sensitive to direct PBH studies.}
\label{compare}
\end{center}
\end{table}
\subsection{Systematic Uncertainties}\label{sec:sys}
\newcommand{\rpm}{\raisebox{.2ex}{$\scriptstyle\pm$}}
The main source of systematic uncertainties within HAWC analyses comes from discrepancies between the data and the simulated Monte Carlo events, which stem from uncertainties in the modeling of the detector. The dominant sources of these discrepancies are PMT efficiency and late light simulation. The PMT efficiency cannot be precisely determined using the calibration system; instead, an event selection based on charge and timing cuts is implemented to identify incident vertical muons. Since vertical muons provide a mono-energetic source of light, they can be used to measure the relative efficiency of each PMT by matching the muon peak position to the expected position from simulations, which is used to measure the range of uncertainties. Uncertainty also arises from mis-modeling of late light in air showers stemming from a discrepancy between the time width of the laser pulse used for calibration and the time structure of actual showers.

Less dominant contributors include PMT thresholds, angular resolution discrepancy, and charge uncertainty. The PMT threshold in simulations may be, based on observations of the cosmic-ray rate, deviating $\pm 0.05~\mathrm{PE}$ from the actual value; the 68\% containment radius of the reconstructed gamma-ray direction around the true direction in the HAWC detector Monte Carlo model is underestimated by $\sim5\%$; and the charge uncertainty stems from relative differences in photon detection efficiency from PMT to PMT, encapsulating how much of a PMT measurement will vary for a fixed amount of light.

\begin{center}
\begin{figure}[htb!]
\centering
\includegraphics[width=0.8\textwidth]{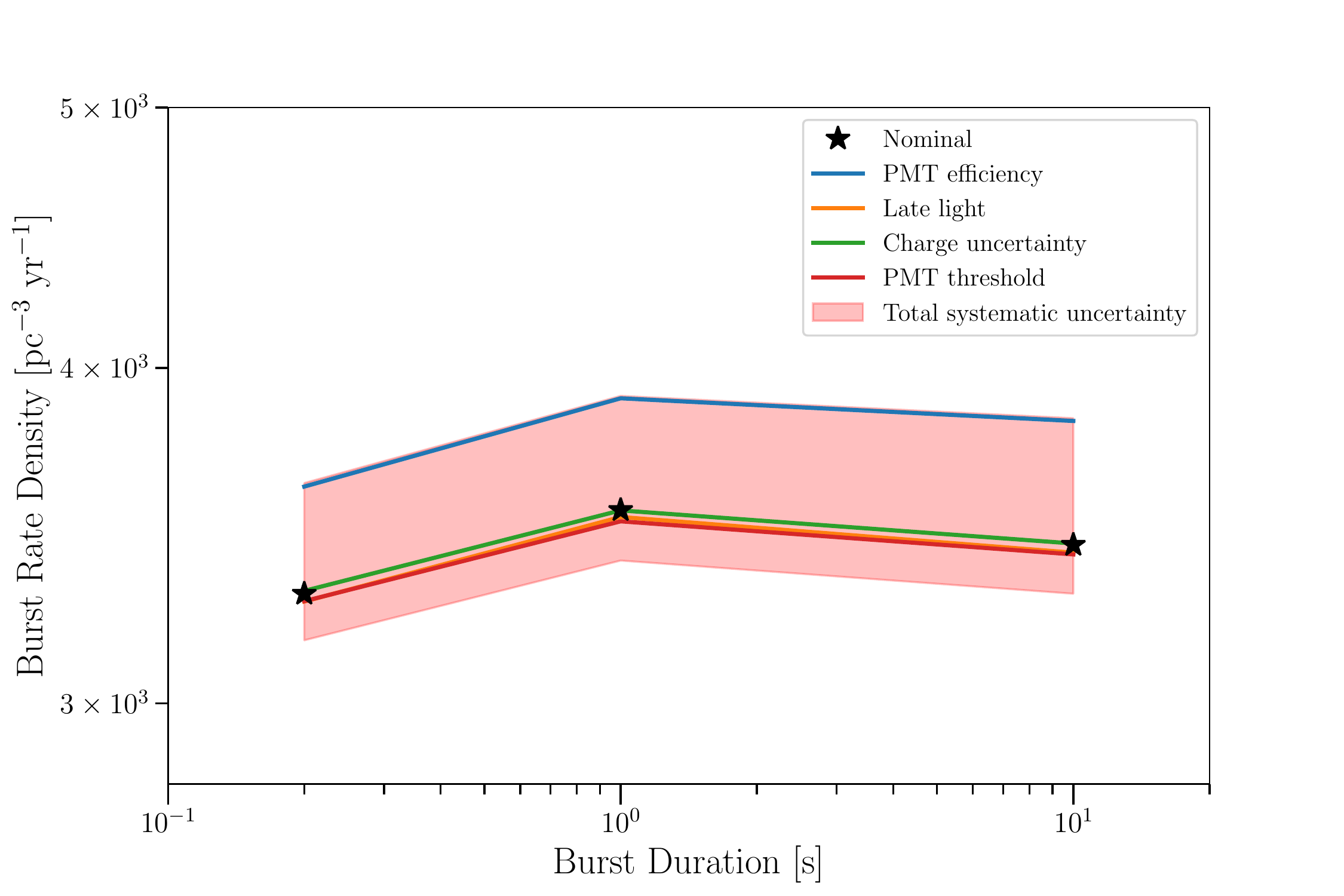}
\caption{The spread in resulting upper limits obtained by fluctuating each source of systematic error to determine its contribution. The black stars are our final results for each remaining lifetime to search, and the thick red band represents the total systematic uncertainty found by adding each source in quadrature.} 
\label{fig:systematics}
\end{figure}
\end{center}

These effects are described in more detail in Ref.~\cite{Abeysekara:2019} and have been evaluated for correlations with none found. To account for these uncertainties in this analysis, we made a new model histogram for each source of uncertainty based off of simulations using different detector models and then repeated the analysis. We assume these variables are independent and add each source of systematic uncertainty in quadrature with the others, resulting in approximately $+10.6\%$, $-4.3\%$ uncertainty in our results (shown in Figure~\ref{fig:systematics}). These uncertainties, being significantly smaller than the $1\sigma$ and $2\sigma$ containment bands of the expected limit, are not shown in Figure~\ref{fig:result}.
\subsection{Conclusions}
We evaluated three years of HAWC transient search data for PBH bursts. No significant signal was found, so we used the log-likelihood method described in Section~\ref{sec:limit} to set upper limits on the local burst rate density at $99\%$ confidence. We set a limit of $\sim3400~\mathrm{pc^{-3}~yr^{-1}}$ using a burst duration of 10~s, the most constraining limit placed to date for very nearby PBHs. Note that the burst duration is a search parameter, not a physical parameter, thus differences between the limits placed for each burst duration searched are due to differences in the signal-to-background ratio within HAWC.

Planned future work to improve this analysis includes a dedicated PBH search working directly with HAWC data rather than results from a previous blind transient search on HAWC data. This would also incorporate more durations to search, energy estimators, and more days of HAWC data. Statistical improvements for such a study would include full likelihood profiles in both time and space, as well as stacking of these likelihoods to ensure the signal is well-defined for any zenith angle. Also, data using the newly-built outrigger array will increase the sensitivity at $>10~\mathrm{TeV}$ \cite{Joshi:2019ycc}, relevant for hard-spectrum sources such as PBHs. We anticipate significant enhancement in our ability to search for PBH bursts with the application of such improvements.  


\acknowledgments

We acknowledge the support from: the US National Science Foundation (NSF); the US Department of Energy Office of High-Energy Physics; the Laboratory Directed Research and Development (LDRD) program of Los Alamos National Laboratory; Consejo Nacional de Ciencia y Tecnolog\'{\'i}a (CONACyT), M{\'e}xico, grants 271051, 232656, 260378, 179588, 254964, 258865, 243290, 132197, A1-S-46288, A1-S-22784, c{\'a}tedras 873, 1563, 341, 323, Red HAWC, M{\'e}xico; DGAPA-UNAM grants AG100317, IN111315, IN111716-3, IN111419, IA102019, IN112218; VIEP-BUAP; PIFI 2012, 2013, PROFOCIE 2014, 2015; FAPESP support No. 2015/15897-1 and 2017/03680-3, and the LNCC/MCTI, Brazil; the University of Wisconsin Alumni Research Foundation; the Institute of Geophysics, Planetary Physics, and Signatures at Los Alamos National Laboratory; Polish Science Centre grant DEC-2018/31/B/ST9/01069, DEC-2017/27/B/ST9/02272; Coordinaci{\'o}n de la Investigaci{\'o}n Cient\'{\'i}fica de la Universidad Michoacana; Royal Society - Newton Advanced Fellowship 180385. Thanks to Scott Delay, Luciano D\'{\'i}az and Eduardo Murrieta for technical support.





\bibliography{Refs}



\end{document}